\documentclass[aps,preprint,showpacs,floats,epsf,epsfig,nofootinbib,12pt]{revtex4}
\usepackage{graphics}
\usepackage{epsfig,epsf,psfrag}
\usepackage{times}
\usepackage{float}
\usepackage{color}
\usepackage{amsfonts,amssymb,stmaryrd,latexsym,amsmath}
\usepackage{multirow}
\usepackage{array}
\usepackage{hhline}
\usepackage{booktabs}
\usepackage{enumerate}
\usepackage{slashed}
\usepackage{subfigure}
\usepackage[colorlinks, citecolor=blue,anchorcolor=red,menucolor=red, linkcolor=red,filecolor=red,runcolor=red,urlcolor=blue,frenchlinks=red]{hyperref}
\usepackage{epstopdf}
\usepackage{braket}

\def\beq{\begin{eqnarray}}
\def\eaeq{\end{eqnarray}}
\def\non{\nonumber}

\def\ra{\rangle}

\begin{document}


\title{Weak decays of the triply heavy baryons in the three-quark picture with the light-front quark model}

\vspace{1cm}

\author{ Fang Lu, Hong-Wei Ke\footnote{khw020056@tju.edu.cn, corresponding author} and Xiao-Hai Liu\footnote{xiaohai.liu@tju.edu.cn}}

\affiliation{ School of Science, Tianjin University, Tianjin 300072, China}

\vspace{12cm}

\begin{abstract}
We investigate the  weak decays of the triply heavy baryon $\Omega_{QQQ}$ in the
light-front quark model. 
 Since $\Omega_{QQQ}$ consists of three indistinguishable identical heavy quarks, the commonly adopted quark-diquark picture does not seem to be valid anymore.
Instead, we employ the three-quark picture for baryons
where the three quarks are regarded as individual quarks.
We calculate the hadronic form factors for the transitions and give predictions for the decay widths of the semi-leptonic decay modes $\Omega_{ccc}\to \Xi_{cc}/ \Omega_{cc}+ l\bar{\nu}_l$,
 $\Omega_{bbb}\to \Xi _{bb}+ l \bar{\nu}_l$
 and the non-leptonic decay modes $\Omega_{ccc}\to \Xi_{cc}/\Omega_{cc}+ M$,  $\Omega_{bbb}\to \Xi_{bb}+ M$.
 Our study can be a guide for future experiments to discover the triply heavy baryons.


\end{abstract}

\maketitle

\section{Introduction}

According to the conventional quark model, baryons consist of three valence quarks and
mesons  are composed of one quark and one antiquark. Compared to the
two-body meson system, the theoretical study of the three-body baryon system is far more complicated, therefore the researches on baryons
are much more behind than those on mesons, especially for the heavy flavor baryons.
Experimentally, higher energy is necessary to produce the heavy baryons, and usually, the production rates are not very large.
In the last century, the properties of the light baryons
have been extensively studied in many models with great success. However, heavy baryon spectroscopy has been hampered
by insufficient experimental data. As experimental techniques have advanced, more and more new data on the heavy baryons have
been accumulated in recent years. So far, the baryons containing one heavy quark ($b$, $c$) have been extensively studied
and experimentally observed. In 2017, the LHCb collaboration observed the doubly charmed baryon $\Xi_{cc}^{++}$ in the final state
$\Lambda_c^{+}K^{-}\pi^+\pi^+$\cite{LHCb:2017iph}, which was subsequently confirmed in decays
$\Xi_{cc}^{++}\to\Xi_{c}^+\pi^+$and $\Xi_{cc}^{++}\to\Xi_{c}^{\prime +}\pi^+$ \cite{LHCb:2018pcs,LHCb:2022rpd}.
This discovery has spurred extensive research on the doubly heavy baryons \cite{Wang:2017mqp,Zhao:2018mrg,Yu:2017zst,Aliev:2022tvs,Hu:2022xzu}.
The triply heavy baryons, consisting of three heavy flavor quarks, have not yet been discovered.
Nevertheless, theoretical studies on the triply heavy baryons have been ongoing, including their
mass spectra \cite{Patel:2008mv,Brown:2014ena,Yang:2019lsg,Chen:2011mb,Faustov:2021qqf,Zheng:2010zzc,Jia:2006gw,Salehi:2022lkt}
and pertinent decay modes \cite{Huang:2021jxt,Wang:2018utj,Wang:2022ias,Zhao:2022vfr,Geng:2017mxn}.
In this work we focus on the the weak decays of the triply heavy baryon $\Omega_{QQQ}$, which is
composed of three identical heavy quarks $Q$ ($b$ or $c$), to the doubly heavy baryon $\mathcal B_{QQ}$.

We employ the light front quark model (LFQM) to investigate the weak decay processes. LFQM is a relativistic quark model which was initially used to study the decays of mesons \cite{ODonnell:1995dio,Belyaev:1997iu,Choi:1998jd,DeWitt:2003nxf,Cheng:2003sm,Ke:2011mu,Chang:2018zjq}
and has been extended to deal with the transitions between two heavy baryons \cite{Ke:2007tg,Ke:2012wa,Ke:2017eqo}.
The most common way of dealing with the baryons is to transform a three-body problem into a two-body one in the
quark-diquark picture ansatz where the spectator quarks are treated as the diquark.
In Refs.~\cite{Ke:2007tg,Ke:2012wa,Ke:2017eqo,Zhao:2018zcb,Guo:2005qa,Hu:2020mxk,Chua:2019lgb,Chua:2018lfa},
the transition between  baryons was studied with the quark-diquark picture
and  the results are in good agreement with the experiment.
In Refs. \cite{Wang:2022ias,Zhao:2022vfr}, the authors employed the quark-diquark picture to
study the weak decays of the triply heavy baryon $\Omega_{QQQ}$ within the LFQM framework.
In general, light quark pair [$q_1q_2$] in the single-heavy baryon is regarded as the diquark
as well as the heavy quark pair [$Q_1Q_2$]  in the doubly heavy baryon.
However, for the triply heavy baryon, as $\Omega_{QQQ}$ consists entirely of heavy quarks,
the interactions between any two quarks are equally strong,
quark-diquark picture does not seem to be so appropriate anymore.
In addition, for the weak decay of the doubly charmed baryon $\Xi_{cc}^{++}$, the diquark in the initial state and that in the final state are not spectators,
so there are also some difficulties in using the quark-diquark picture here \cite{Ke:2019lcf}. 
As discussed in Refs.~\cite{Ke:2019smy,Ke:2021pxk,Li:2021qod,Li:2021kfb,Li:2022hcn,Geng:2020gjh,Lu:2023rmq}, the recipe is to use the three-quark picture, which can compensate the shortcoming of the quark-diquark picture.
In this work, we also employ the three-quark picture and use the same approach as that in the Ref.~\cite{Lu:2023rmq} to explore the decays of $\Omega_{QQQ}$ within the LFQM framework.



This paper is organized as follows:  In
section II we present the vertex functions of the heavy flavor baryons,
and  derive the form factors of the transitions $\Omega_{ccc}\to \Xi_{cc}/\Omega_{cc}$
and $\Omega_{bbb}\to \Xi_{bb}$  in the LFQM; In section III we present numerical
results for the transitions $\Omega_{ccc}\to \Xi_{cc}/\Omega_{cc}$
and $\Omega_{bbb}\to\Xi_{bb}$ along with all
necessary input parameters, and then we calculate the form factors and the decay widths of related semi-leptonic and non-leptonic decays;
Section IV is devoted to the conclusions and discussions.

\section{ $\Omega_{QQQ}\to \mathcal B_{QQ}$  in LFQM}
\subsection{The vertex functions}

In Ref. \cite{Lu:2023rmq}, under the three-quark picture, the vertex function of a baryon
$\mathcal B_{QQ}$ ($\Xi_{cc}$, $ \Omega_{cc}$ or $\Xi_{bb}$) with the total spin $S=1/2$ and total momentum $P$ is
\begin{eqnarray}\label{baryonv1}
  && |\mathcal B_{QQ}(P,S,S_z)\rangle=\int\{d^3\tilde p_1\}\{d^3\tilde
p_2\}\{d^3\tilde p_3\} \,
  2(2\pi)^3\delta^3(\tilde{P}-\tilde{p_1}-\tilde{p_2}-\tilde{p_3}) \nonumber\\
 &&\times\sum_{\lambda_1,\lambda_2,\lambda_3}\Psi_{\mathcal B_{QQ}}^{SS_z}(\tilde{p}_1,\tilde{p}_2,\tilde{p}_3,\lambda_1,\lambda_2,\lambda_3)
  \mathcal{C}^{\alpha\beta\gamma}\mathcal{F}_{QQq} \left|\right. Q_{\alpha}(p_1,\lambda_1)Q_{\beta}(p_2,\lambda_2)q_{ \gamma}(p_3,\lambda_3)\rangle,
  \end{eqnarray}
where $q$  denotes light quark ($u$, $d$ or $s$),  $\lambda_i$ and $p_i\, (i=1,2,3)$ are helicities and
light-front momenta of the quarks,
$\mathcal{C}^{\alpha\beta\gamma}$ and $\mathcal{F}_{QQq}$
are the color and flavor wave functions.

The vertex function of $\Omega_{QQQ}$ with the total spin $S=3/2$ and total momentum $P$  is
\begin{eqnarray}\label{baryonv2}
  && |\Omega_{QQQ} (P,S,S_z)\rangle=\int\{d^3\tilde p_1\}\{d^3\tilde
p_2\}\{d^3\tilde p_3\} \,
  2(2\pi)^3\delta^3(\tilde{P}-\tilde{p_1}-\tilde{p_2}-\tilde{p_3}) \nonumber\\
 &&\times\sum_{\lambda_1,\lambda_2,\lambda_3}\Psi_{\Omega_{QQQ}}^{SS_z}(\tilde{p}_1,\tilde{p}_2,\tilde{p}_3,\lambda_1,\lambda_2,\lambda_3)
  \mathcal{C}^{\alpha\beta\gamma}\mathcal{F}_{QQQ} \left|\right. Q_{\alpha}(p_1,\lambda_1)Q_{\beta}(p_2,\lambda_2)Q_{\gamma}(p_3,\lambda_3)\rangle.
  \end{eqnarray}
 where $Q$  denotes heavy quark ($b$ or $c$).

  In the light-front quark model, the on-mass-shell light-front momentum $p$ is defined as
\begin{equation}
\tilde{p}=(p^{+},p_{\perp}),  \quad p_{\perp} =(p^{1},p^{2}), \quad p^{-} =\frac{m^2+p^2_{\perp}}{p^{+}} \ , \quad
\{d^3p\} = \frac{dp^{+}d^2p_{\perp}}{2(2\pi)^3} \ .
\end{equation}
Internal variables $(x_i, k_{i\perp})$ are introduced  to facilitate the calculation  (
$i=1,2,3$) and the kinematics constituents are
\begin{eqnarray}
&&p^+_i=x_i P^+, \qquad p_{i\perp}=x_i P_{\perp}+k_{i\perp},
 \qquad x_1+x_2+x_3=1, \qquad k_{1\perp}+k_{2\perp}+k_{3\perp}=0,
\end{eqnarray}
where $x_i$  is the momentum fraction and satisfies the relation $0<x_1, x_2, x_3<1$.

  The momentum-space wave functions $\Psi_{\Omega_{QQQ}}^{SS_z}$ and $\Psi_{\mathcal B_{QQ}}^{SS_z}$ \cite{Lu:2023rmq} are
\begin{eqnarray}\label{wave_1}
\Psi_{\Omega_{QQQ}}^{SS_z}(\tilde{p}_i,\lambda_i)=&&A_0 \bar
u(p_3,\lambda_3)[(\bar
P\!\!\!\!\slash+M_0) \gamma_{\perp}^\alpha]v(p_2,\lambda_2)\bar
u(p_1,\lambda_1) u_{\alpha}(\bar
P,S)\phi(x_i,k_{i\perp}),\nonumber\\
\Psi_{\mathcal B_{QQ}}^{SS_z}(\tilde{p}_i,\lambda_i)=&&A_1 \bar
u(p_3,\lambda_3)[(\bar
P\!\!\!\!\slash+M_0) \gamma_{\perp}^\beta]v(p_2,\lambda_2)\bar
u(p_1,\lambda_1) \gamma_{\perp\beta}\gamma_{5} u(\bar P,S) \phi(x_i,k_{i\perp}),
\end{eqnarray}
where the $u$, $ \bar v$ and $u_{\alpha}$ are spinors, $p_i (i=1,2,3)$ is the
momentum of the  constituent quark, $\lambda_i (i=1,2,3)$ is the helicity of the  constituent quark,
$\bar P$ ($\bar P=p_1+p_2+p_3 $) is the sum of the momenta of the  constituent quarks,
$M_0$ is the invariant mass of the baryon
and $\gamma_{\perp}^\alpha=\gamma^{\alpha}-v\!\!\!\slash v^{\alpha}$.

The factors $A_0$ and $A_1$ in Eq.~(\ref{wave_1}) read
\begin{eqnarray}
A_0
&&=\frac{1}{4\sqrt{2P^+M_0^3(m_1+e_1)(m_2+e_2)(m_3+e_3)}},\nonumber\\
A_1&&=\frac{1}{4\sqrt{3P^+M_0^3(m_1+e_1)(m_2+e_2)(m_3+e_3)}}.
\end{eqnarray}
The invariant mass square $M_0^2$ is
defined as a function of the internal variables $x_i$ and $k_{i\perp}$
 \begin{eqnarray} \label{eq:Mpz}
  M_0^2=\frac{k_{1\perp}^2+m_1^2}{x_1}+
        \frac{k_{2\perp}^2+m_2^2}{x_2}+\frac{k_{3\perp}^2+m_3^2}{x_3},
 \end{eqnarray}
with the internal momentum
 \begin{eqnarray}
 k_i=(k_i^-,k_i^+,k_{i\bot})=(e_i-k_{iz},e_i+k_{iz},k_{i\bot})=
  (\frac{m_i^2+k_{i\bot}^2}{x_iM_0},x_iM_0,k_{i\bot}),
 \end{eqnarray}
 and it is easy to obtain
 \begin{eqnarray}
  e_i&=&\frac{x_iM_0}{2}+\frac{m_i^2+k_{i\perp}^2}{2x_iM_0}
 ,\non\\
 k_{iz}&=&\frac{x_iM_0}{2}-\frac{m_i^2+k_{i\perp}^2}{2x_iM_0},
 \label{e_i}
 \end{eqnarray}
{where $e_i$ is the energy of the $i$-th constituent, and
they obey the condition $e_1+e_2+e_3=M_0$. The transverse $k_{i\bot}$ and $z$ direction $k_{iz}$ components  constitute a momentum vector $\vec k_i=(k_{i\bot}, k_{iz})$.}

The spatial wave function  $\phi(x_i, k_{i\perp})$ in Eq.~(\ref{wave_1}) is defined as
 \begin{eqnarray}\label{A122}
\phi(x_1,x_2,x_3,k_{1\perp},k_{2\perp},k_{3\perp})=\sqrt{\frac{e_1e_2e_3}{x_1x_2x_3M_0}}
\varphi(\overrightarrow{k}_1,\beta_1)\varphi(\frac{\overrightarrow{k}_2-\overrightarrow{k}_3}{2},\beta_{23})
 \end{eqnarray}
where
$\varphi(\overrightarrow{k},\beta)=4(\frac{\pi}{\beta^2})^{3/4}{\rm
exp}(\frac{-k_z^2-k^2_\perp}{2\beta^2})$ is the phenomenological Gaussian form and
$\beta$ is a non-perturbative parameter that describes the inner structure of baryon~\cite{Cheng:2004ew,Cheng:2004cc}.

\subsection{The form factors of $\Omega_{QQQ}\to \mathcal B_{QQ}$}

\begin{figure}
\begin{center}
\scalebox{0.7}{\includegraphics{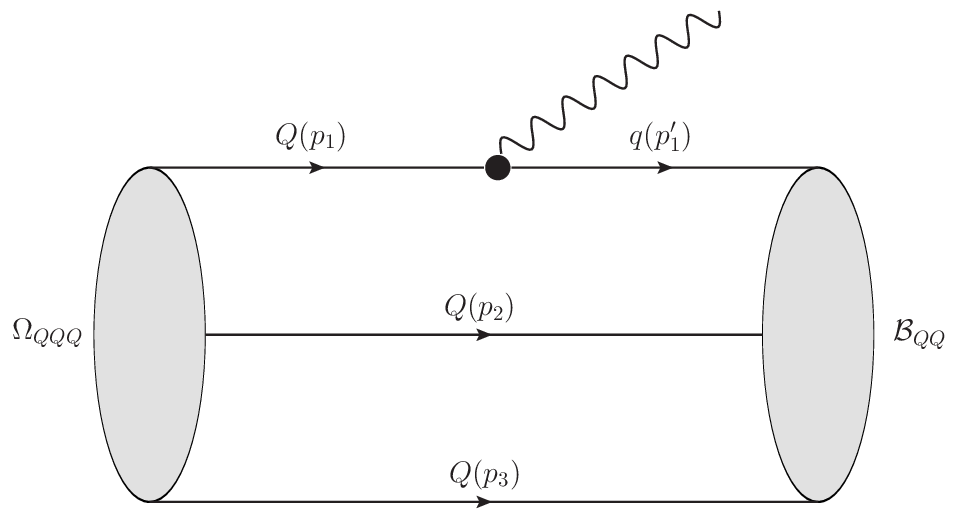}}
\end{center}
\caption{The schematic diagram for the triply heavy baryon decaying into the doubly heavy baryon $\Omega_{QQQ}\to \mathcal{B}_{QQ}$. The black dot denotes the $V-A$ weak current vertex. } \label{t1}
\end{figure}

The lowest order Feynman diagram responsible for the weak decay $\Omega_{QQQ}\to \mathcal B_{QQ}$
is shown in Fig.\ref{t1}. Following the approach given in
Refs. \cite{Ke:2007tg,Ke:2019smy,Lu:2023rmq},
the transition matrix element can be calculated with the vertex
functions  $|\Omega_{QQQ}(P,3/2,S_z)$$ (\mid
\Omega_{QQQ}\ra$) and $|\mathcal B_{QQ}(P',1/2,S'_z)\rangle$ ($\mid \mathcal B_{QQ} \ra$). The matrix element takes the form

\begin{eqnarray}\label{s1}
&& \langle\mathcal B_{QQ}\mid
\bar{q}
\gamma^{\mu} (1-\gamma_{5}) Q \mid\Omega_{QQQ}  \ra = \nonumber \\
 && \int\frac{\{d^3 \tilde p_2\}\{d^3 \tilde p_3\}\phi^*(x^\prime_i,k^\prime_{i \perp})
  \phi(x_i,k_{i \perp}){\rm Tr}[\gamma_{\perp}^\beta(\bar{P^\prime}\!\!\!\!\!\slash+M_0^\prime)(p_3\!\!\!\!\!\slash+m_3)
  (\bar{P}\!\!\!\!\slash+M_0)\gamma_{\perp}^\alpha(p_2\!\!\!\!\!\slash-m_2)]}{16\sqrt{6p^+_1p^{\prime+}_1{P}^+P^{\prime+}M_0^3M_0^{\prime 3}(m_1+e_1)
 (m_2+e_2)(m_3+e_3)(m_1^\prime+e_1^\prime)
 (m_2^\prime+e_2^\prime)(m_3^\prime+e_3^\prime)}}\nonumber \\
  &&\times  \bar{u}(\bar{P}^\prime,S^\prime_z)\gamma_{\perp\beta}\gamma_{5}
  (p_1\!\!\!\!\!\slash ^\prime +m^\prime_1)\gamma^{\mu}(1-\gamma_{5})
  (p_1\!\!\!\!\!\slash+m_1) u_{\alpha}(\bar{P},S_z),
\end{eqnarray}
where
\begin{eqnarray}
&&m_1=m_Q, \qquad m_1^{\prime}=m_{q}, \qquad m_2=m_Q, \qquad m_3=m_Q, \cr
&& \gamma_{\perp}^\alpha=\gamma^{\alpha}-v\!\!\!\slash v^{\alpha},
\qquad \gamma_{\perp}^\beta=\gamma^{\beta}-v^\prime\!\!\!\!\slash v^{\prime\beta}
\end{eqnarray}
with $v$ ($P/M$) and $v^\prime$ ($P^\prime/M^\prime$) being the velocities of the initial state $\Omega_{QQQ}$ and
final state $\mathcal B_{QQ}$, respectively. $M(M^\prime)$ is the mass of $\Omega_{QQQ}$ ($\mathcal B_{QQ}$)
and $P (P^\prime)$ is the four-momentum of $\Omega_{QQQ}$ ($\mathcal B_{QQ}$).

Setting $\tilde{p}_1-\tilde{p}^\prime_1=\tilde{q}^\prime$, $\tilde{p}^\prime_2=\tilde{p}_2$,  $\tilde{p}_3=\tilde{p}^\prime_3$ ,
we can get
\begin{eqnarray}
&&x^\prime_{1,2,3}=x_{1,2,3},\quad k^\prime_{1\perp}=k_{1\perp} - (1-x_1) q^\prime_{\perp},\quad k^\prime_{2\perp}=k_{2\perp} + x_2 q^\prime_{\perp},
\quad k^\prime_{3\perp}=k_{3\perp} + x_3 q^\prime_{\perp},
\end{eqnarray}
where $q^\prime$ denotes the transfer momentum $q^\prime\equiv P-P^\prime$ and $p_i^{(\prime)2}=m_i^{(\prime)2}$.

The form factors for the transition $\Omega_{QQQ}\rightarrow\mathcal B_{QQ}$ $(3/2^+\rightarrow 1/2^+)$  are defined as
\begin{eqnarray}\label{s2}
 \langle\mathcal B_{QQ}\mid
\bar{q}
\gamma^{\mu} (1-\gamma_{5}) Q \mid\Omega_{QQQ}\ra =  \bar{u}(P',S'_z) \Bigg[ \gamma^{\mu} P'^\alpha \frac{f_{1}(q'^{2})}{M}+ \frac{f_{2}(q'^{2})}{M(M-M')} P'^\alpha P^\mu
\nonumber\\+\frac{f_{3}(q'^{2})}{M(M-M')} P'^\alpha P'^\mu+f_4(q'^{2})g^{\alpha \mu}
 \Bigg]\gamma_{5} u_{\alpha}(P,S_z)-\bar u(P',S'_z) \Bigg[\gamma^{\mu} P'^\alpha \frac{g_{1}(q'^{2})}{M} \nonumber \\
 + \frac{g_{2}(q'^{2})}{M(M-M')} P'^\alpha P^\mu
 +\frac{g_{3}(q'^{2})}{M(M-M')} P'^\alpha P'^\mu+g_4(q'^{2})g^{\alpha \mu}
 \Bigg] u_{\alpha}(P,S_z).
\end{eqnarray}
The momentum $P$ $(P')$ satisfies the relation $P^{(\prime)2}=M^{(\prime)2}$, while the
$\bar P$ ($\bar P^\prime$) is the sum of the momenta of the constituent quarks
and does not obey the  on-shell condition. Nonetheless, we can still take the approximation
$P^{(\prime)}\simeq\bar P^{(\prime)}$.

In order to get the form factors, we can multiply the following terms $\bar {u}_{\xi}(\bar
P,S_z)\gamma_{\mu}\bar P'^\xi \gamma_{5} u(\bar P',S'_z)$, $\bar
{u}_{\xi}(\bar P,S_z)\bar P'_{\mu}\bar P'^\xi \gamma_{5} u \bar P',S'_z)$ and
$\bar {u}_{\xi}(\bar P,S_z) \bar P_{\mu}\bar P'^\xi \gamma_{5} u(\bar
P',S'_z)$, $\bar {u}_{\xi}(\bar P,S_z) g_{\mu}^{\xi} \gamma_{5} u(\bar
P',S'_z)$ to the right sides of both Eq. (\ref{s1}) and Eq. (\ref{s2}) and then summing over
the polarizations of all states. Then we have four algebraic equations, each of which contains the form factors $f_{1}$, $f_{2}$, $f_{3}$ and $f_{4}$. Solving these equations, we obtain the explicit expressions of the form factors  $f_{i}$($i=1,2,3,4$) (See Appendix $A$ for details).

  Similarly, multiplying the expressions  $\bar {u}_{\xi}(\bar P,S_z)
\gamma_{\mu}\bar P'^\xi  u(\bar P',S'_z)$ , $\bar{u}_{\xi}
(\bar P,S_z)\bar P'_{\mu}\bar P'^\xi  u \bar P',S'_z)$,
$\bar {u}_{\xi}(\bar P,S_z) \bar P_{\mu}\bar P'^\xi u(\bar
P',S'_z)$, $\bar {u}_{\xi}(\bar P,S_z) g_{\mu}^{\xi} u(\bar
P',S'_z)$ to the right sides of both Eq. (\ref{s1}) and Eq. (\ref{s2}), we can get
form factors  $g_{i}$($i=1,2,3,4$).

The flavor wave function of $\Omega_{QQQ}$ is
\begin{eqnarray}
\Omega_{QQQ}
&=&\frac{1}{\sqrt{3}}([Q_1Q_2]_AQ_3+[Q_1Q_3]_AQ_2+[Q_2Q_3]_AQ_1).
\end{eqnarray}
Because the baryon  $\Omega_{QQQ}$ is composed entirely of heavy quark ($c$ or $b$), there is an overlap factor $\sqrt 3$.

\section{Numerical Results}
\label{section_numerical}

\begin{table}
\caption{The  masses of the involved quarks and baryons (in units of
 GeV).}\label{Tab:t1}
\begin{ruledtabular}
\begin{tabular}{cccccccccc}
  $m_c$ &$m_b$ & $m_{d}$  & $m_{u}$ & $m_s$ &$\Xi_{cc}$\cite{ParticleDataGroup:2022pth}   &$\Omega_{cc}$ \cite{Brown:2014ena} &$\Xi_{bb}$ \cite{Brown:2014ena}  &$\Omega_{ccc}$ \cite{Faustov:2021qqf}  &$\Omega_{bbb}$ \cite{Faustov:2021qqf} \\\hline
  $1.3$   &4.4   & 0.25      &0.25       &0.5             &3.621              &3.738     &10.14      &4.712   &14.47
\end{tabular}
\end{ruledtabular}
\end{table}

\begin{table}
\caption{The values of the parameter $\beta$ (in units of GeV).}\label{Tab:t11}
\begin{ruledtabular}
\begin{tabular}{ccccc}
  $\beta_{c\bar c}$  &$\beta_{b\bar b}$  &$\beta_{c\bar d}$ &$\beta_{c\bar s}$ &$\beta_{b\bar u}$ \\\hline
    0.6545  &1.391   &0.4641     &0.5375   &0.5479
\end{tabular}
\end{ruledtabular}
\end{table}

\subsection{The $\Omega_{QQQ}\to \mathcal B_{QQ}$  form factors  }
In order to calculate the form factors numerically, firstly it is necessary to determine the relevant parameters in the model.
We adopt the quark mass values given in Ref. \cite{Cheng:2003sm}. For the mass of $\Xi_{cc}$, the experimental value is adopted~\cite{ParticleDataGroup:2022pth}. For the other heavy baryons, we adopt the values given in Refs. \cite{Brown:2014ena,Faustov:2021qqf}. The explicit parameter values are listed in Table \ref{Tab:t1}.

The reciprocal of $\beta$ is generally related to the charge radius of two constituents. The rule for selecting $\beta$ is derived from Ref. \cite{Ke:2019lcf,Lu:2023rmq}. Since $\Omega_{QQQ}$ is composed entirely of heavy quark $Q$, the distances between any two quarks are equal.
We choose $\beta_1=\beta_{Q[QQ]}=\frac{2}{\sqrt 3}\beta_{QQ}$ in the initial state. For the doubly heavy baryon of the final state, we set $\beta_{d[cc]}=\sqrt{2}\beta_{c\bar d}$, $\beta_{s[cc]}=\sqrt{2}\beta_{c\bar s}$.
However, since we know little about the structure of the triply heavy baryons, we make predictions with different $\beta_{QQ}$:

(1) Case I : Assuming that the spectators $QQ$ in the initial state remains unchanged in the final state, which implies the distance between quarks is unchanged, then we set $\beta_{QQ}=2\beta_{Q\bar Q}$  for both the initial and final states.

(2) Case II: Assuming that the distance between the two spectators in the initial state is normal but two $Q$ quarks approach each other to form a diquark during the decay process, then we set $\beta_{QQ}=\sqrt{2} \beta_{Q\bar Q}$ in the initial state and $\beta_{QQ}=2\beta_{Q\bar Q}$ in the final state.

The values of the parameter $\beta$ for different states are taken from Ref. \cite{Chang:2018zjq} and listed in Table \ref{Tab:t11}.
With these parameters we calculate the form factors and make theoretical predictions about the transition rates. Anyway, the $\beta$ value is still model dependent, we hope in the future by comparing the theoretical predictions with experimental data, the parameter value can be determined better.

\begin{figure}[hhh]
\begin{center}
\scalebox{0.8}{\includegraphics{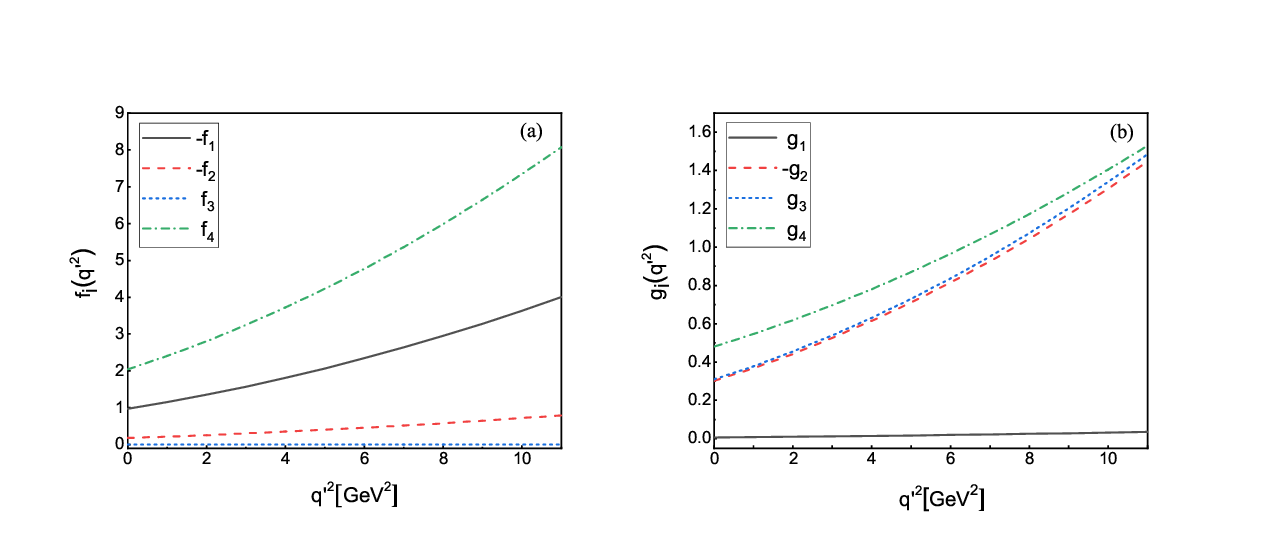}}
\end{center}
\caption{(a)  The form factors  $f_i\; (i=1,2,3,4)$ and (b) the
form factors $g_i\; (i=1,2,3,4)$ of $\Omega_{ccc}\to \Xi_{cc}$
in case I. } \label{f63}
\end{figure}
\begin{figure}[hhh]
\begin{center}
\scalebox{0.8}{\includegraphics{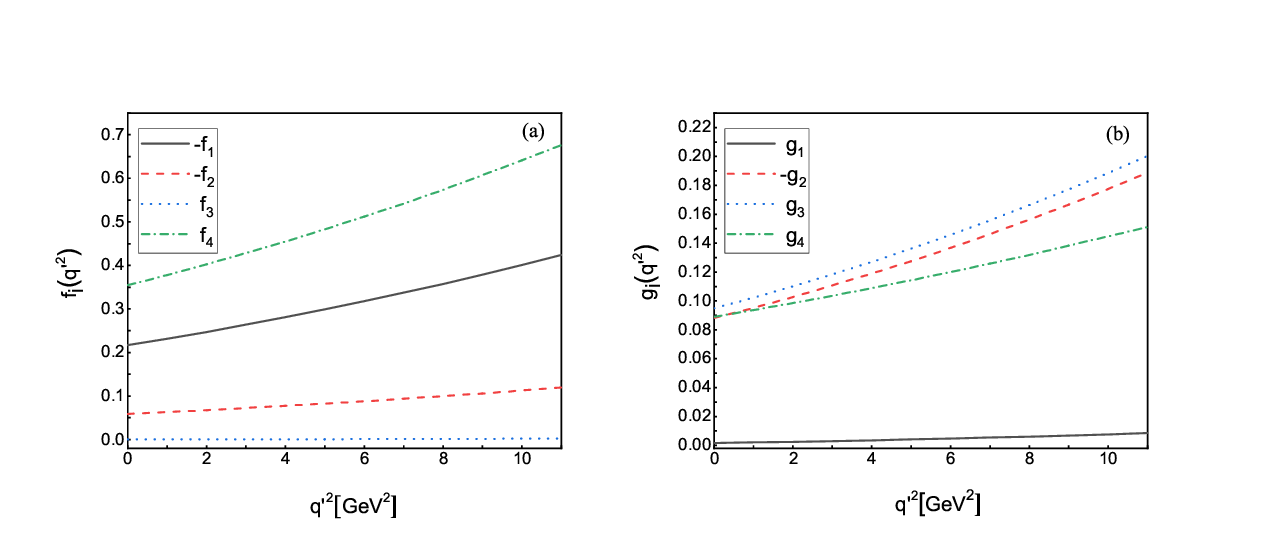}}
\end{center}
\caption{(a)  The form factors  $f_i\; (i=1,2,3,4)$ and (b) the
form factors $g_i\; (i=1,2,3,4)$ of $\Omega_{bbb}\to \Xi_{bb}$ in case I. } \label{f64}
\end{figure}

The  form factors
$f_{i}$ ($i=1,2,3,4$) and $g_{i}$ ($i=1,2,3,4$) is calculated in the frame $q^{\prime +}=0$, i.e., in the space-like region $q^{\prime 2}=-q^{\prime 2}_{\perp}\leq 0$, we can use the same polynomial as that in Ref.  \cite{Lu:2023rmq} to
extend the factors into the time-like region. The form factor takes the form
\begin{eqnarray}\label{s145p}
 F(q'^2)=F(0)+a\frac{q'^2}{M_{\Omega_{QQQ}}^2}+b\left(\frac{q'^2}{M_{\Omega_{QQQ}}^2}\right)^2,
 \end{eqnarray}
and the $F(q'^2)$ represents any of the form factors $f_i$ and $g_i$.
Using the form factors calculated numerically in the space-like region we fit the parameters $a, b $ and $F(0)$ in the unphysical region and then extrapolate  to the physical region with $q'^2\geq 0$ through Eq. (\ref{s145p}). The fitted values of $a,~b$ and $F(0)$ for the form
factors $f_{i}$ and $g_{i}$ are presented in Table \ref{Tab:t2} and \ref{Tab:t4}
for two cases.
The dependence of form factors on $q'^2$ for Case I is depicted in Figs. \ref{f63} and \ref{f64}.
Due to the similarity in the shape of the curves we leave out the graphs for Case II for brevity.

From Figs.~\ref{f63} and \ref{f64}, one can find that the
absolute values of the form factors $f_3(q'^2)$ and $g_1(q'^2)$ are
close to 0 and that of $f_2(q'^2)$ changes slowly.
The curves  of $g_2(q'^2)$ and $g_3(q'^2)$ are close to each other.

\begin{table}
\caption{The $\Omega_{ccc}\to \Xi_{cc}$ form factors given in the
 polynomial form (Case I in the left side and Case II in the right side).}\label{Tab:t2}
\begin{ruledtabular}
\begin{tabular}{c|ccc|ccc}
  $F$    &  $F(0)$ &  $a$  &  $b$  &  $F(0)$ &  $a$  &  $b$\\\hline
  $f_1$  &  -0.969    &  -3.84   & -4.60 &-1.19 &-5.59 &-7.16 \\
$f_2$    &   -0.182    &   -0.765   & -0.941 &-0.242 &-1.17 &-1.51 \\
  $f_3$  &   0.000967   &  0.00607   &  0.00988  &0.000345 &0.00282 &0.00477\\
  $f_4$  &    2.04   &   7.69   &  9.04 &2.34 &10.5 &13.2 \\
  $g_1$  & 0.00639    & 0.0348    &  0.0461 &0.00334 &0.0231 &0.0339 \\
  $g_2$  &  -0.301   &  -1.41   &  -1.82 &-0.434 &-2.26 & -3.05 \\
 $g_3$  &    0.309   & 1.45    & 1.86 & 0.385 &2.05 &2.74 \\
  $g_4$  &   0.481   &1.40   &1.44 &0.495 &1.76 &1.95
\end{tabular}
\end{ruledtabular}
\end{table}

\begin{table}
\caption{The $\Omega_{bbb}\to \Xi_{bb}$ form factors given in the
  polynomial form (Case I in the left side and Case II in the right side).}\label{Tab:t4}
\begin{ruledtabular}
\begin{tabular}{c|ccc|ccc}
  $F$    &  $F(0)$ &  $a$  &  $b$  &  $F(0)$ &  $a$  &  $b$\\\hline
  $f_1$  &  -0.217    &  -3.01   & -17.6  &-0.271 &-5.023 &-35.2 \\
$f_2$    &   -0.0588   &   -0.872   & -5.35 &-0.0767 &-1.48 &-10.6 \\
  $f_3$  &   0.000279   &  0.0199   & 0.282 &0.0003 &0.0195 &0.271 \\
  $f_4$  &    0.355  &   4.71  & 26.8 & 0.433 &7.79 &53.5 \\
  $g_1$  &  0.00160    & 0.0806   &  0.965 &0.00132 &0.0757 &1.00 \\
  $g_2$  &   -0.0883   &  -1.42  &  -9.33 &-0.128 &-2.62 &-19.7 \\
 $g_3$  &   0.0948 & 1.50    & 9.68 &0.122 &2.49 &18.6\\
  $g_4$  &   0.089   & 0.952  & 4.43 &0.093 &1.39 &8.14
\end{tabular}
\end{ruledtabular}
\end{table}

\subsection{Semi-leptonic decays of $\Omega_{QQQ}\to \mathcal B_{QQ}+ l\bar{\nu}_l$}

\begin{figure}[hhh]
\begin{center}
\scalebox{0.8}{\includegraphics{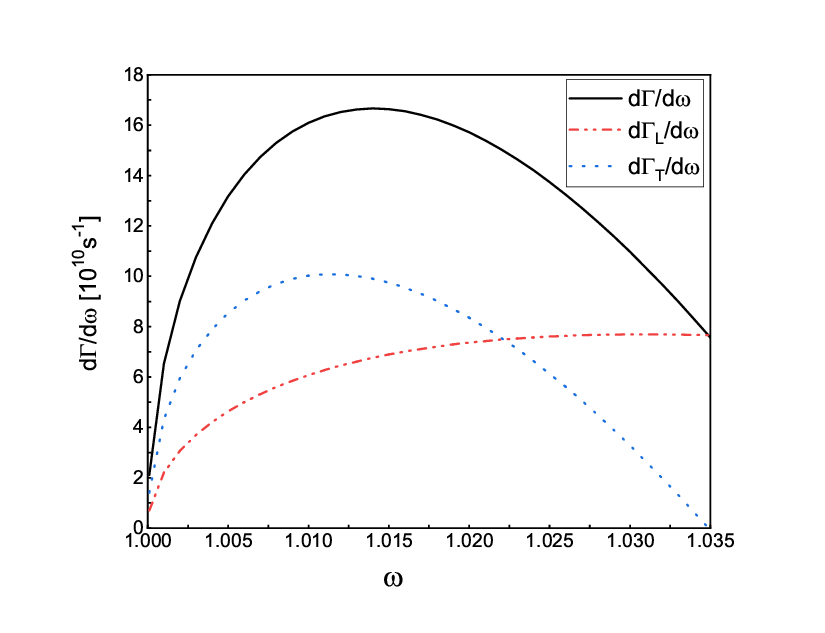}}
\end{center}
\caption{ Differential decay rates $d\Gamma/d\omega$ for the decay
$\Omega_{ccc}\to\Xi_{cc}$ in case I.} \label{f54}
\end{figure}

\begin{figure}[hhh]
\begin{center}
\scalebox{0.8}{\includegraphics{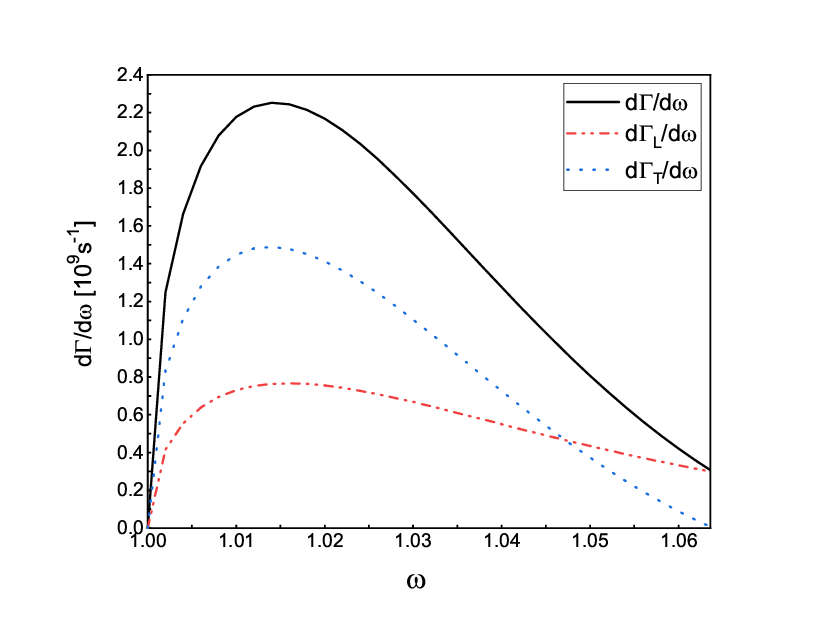}}
\end{center}
\caption{ Differential decay rates $d\Gamma/d\omega$ for the decay
$\Omega_{bbb}\to \Xi_{bb}l\bar{\nu}_l$ in case I.} \label{f56}
\end{figure}

In terms of the form factors obtained in the last subsection, we can calculate the decay rates of $\Omega_{ccc}\to\Xi_{cc}/\Omega_{cc}  l\bar{\nu}_l$
and $\Omega_{bbb}\to \Xi_{bb}l\bar{\nu}_l$ in two cases. The explicit amplitudes are shown in Appendix B.  We also
evaluate the total decay widths and the ratio of the longitudinal to transverse decay rates $R$.
The results are listed in Table \ref{Tab:t5}. In Figs.~\ref{f54} and \ref{f56}, we also plot the $\omega$-dependence of the differential decay rates for
$\Omega_{ccc}\to\Xi_{cc}l\bar{\nu}_l$ and
$\Omega_{bbb}\to \Xi_{bb}l\bar{\nu}_l$ in Case I. We ignore the graph about the differential decay rates depending on $\omega$ of $\Omega_{ccc}\to\Omega_{cc} l\bar{\nu}_l$ since it is similar to Fig.~\ref{f54}, except that its peak value is about 23 times higher than that in Fig.~\ref{f54}.

Our predictions on the ratios of longitudinal to transverse
decay widths $R$ for $\Omega_{ccc}\to\Xi_{cc} l\bar{\nu}_l$
and $\Omega_{ccc}\to\Omega_{cc}  l\bar{\nu}_l$ are close to 1, which are different from the results in Refs.~\cite{Wang:2022ias,Zhao:2022vfr}. However, the $R$ value  for
$\Omega_{bbb}\to\Xi_{bb}  l\bar{\nu}_l$ is close to that in Ref. \cite{Zhao:2022vfr} in two cases.
The difference in $R$ of $\Omega_{ccc}\to  \mathcal B_{cc}l\bar{\nu}_l$ and $\Omega_{bbb}\to\Xi_{bb}  l\bar{\nu}_l$ in our results is due to the mass difference of the initial and final states.
Our predictions on $\Gamma (\Omega_{ccc}\to\Xi_{cc}l\bar{\nu}_l)$ and $\Gamma (\Omega_{ccc}\to\Omega_{cc} l\bar{\nu}_l)$ for Case I
are close to those in  Ref. \cite{Wang:2022ias} where the spectator quark pairs $cc$ are regarded as the diquark.
Their assumption is consistent with ours in Case I.
However, $\Gamma (\Omega_{bbb}\to\Xi_{bb}l\bar{\nu}_l)$ for Case I is an order of magnitude smaller than that in Ref. \cite{Zhao:2022vfr}.
The main reason comes from the difference of the physical pictures and the parameters.
The decay width $\Gamma (\Omega_{bbb}\to\Xi_{bb}l\bar{\nu}_l)$ for Case I is larger than that for  Case II.


\begin{table}
\caption{The widths (in unit $10^{10} s^{-1}$) of  $\Omega_{QQQ}\to \mathcal B_{QQ}+ l\bar{\nu}_l$.}\label{Tab:t5}
\begin{ruledtabular}
\begin{tabular}{c|cc|cc|cc|cc}
    &  \multicolumn{2}{c|}{Case I }
    &  \multicolumn{2}{c|} { Case II }
    &\multicolumn{2}{c|}{\cite{Wang:2022ias}}
    &\multicolumn{2}{c}{\cite{Zhao:2022vfr}}\\\hline
Mode&   $\Gamma$ & $R$ &$\Gamma$  &$R$ & $\Gamma$  & $R$  &$\Gamma$ &$R$ \\\hline
 $\Omega_{ccc}\to\Xi_{cc}l\bar{\nu}_l$  &0.467   & 0.93    &0.546 & 1 &0.484 & 0.64 &4.73 & 0.63 \\\hline
 $\Omega_{ccc}\to\Omega_{cc}  l\bar{\nu}_l$ &  7.27   & 0.98      & 8.37  & 1.04  & 7.52  & 0.90 & 63.1 & 0.67   \\\hline
 $\Omega_{bbb}\to \Xi_{bb}l\bar{\nu}_l$ & $9.17\times 10^{-3}$   &0.66 & $1.54\times 10^{-2}$ &0.66 &-  & - &$8.04\times 10^{-2}$ & 0.77
\end{tabular}
\end{ruledtabular}
\end{table}

\subsection{Non-leptonic decays of $\Omega_{QQQ}\to \mathcal B_{QQ} + M $ }

For the non-leptonic decay modes, we adopt the theoretical framework of factorization assumption
which ignores the interactions between hadrons, so the hadronic matrix element can be written as the product of two independent matrix elements. It is referred to Ref. \cite{Lu:2023rmq}  for more details.

In Table~\ref{Tab:t6}, we show the results of $\Omega_{ccc}\to \Xi_{cc}+ M$ ($\pi, K$), $\Omega_{ccc}\to \Omega_{cc}+ M$ ($\pi, K$)
and $\Omega_{bbb}\to \Xi_{bb} + M$ ($\pi, K, D, D_{s}$) for two cases.
From Table \ref{Tab:t6}, one may notice that the results for Case II is about 1.5 times larger than those for Case I. Our predictions on $\Gamma ( \Omega_{bbb}\to \Xi_{bb}+M)$ in Case I are about 1.5 times as large as those in Ref.~\cite{Zhao:2022vfr}, while  those on $\Gamma ( \Omega_{ccc}\to \Xi_{cc}+M)$  and $\Gamma ( \Omega_{ccc}\to \Omega_{cc}+ M)$ are about half as large. Besides, our results on $\Gamma ( \Omega_{ccc}\to \Omega_{cc}+ M)$ for Case I are close to those in \cite{Wang:2022ias}, but the decay widths on $ \Omega_{ccc}\to \Xi_{cc}+M$ are twice as large as those in \cite{Wang:2022ias}.


\begin{table}
\caption{The widths (in unit $10^{10} s^{-1}$) of $\Omega_{QQQ}\to \mathcal B_{QQ} + M $.}\label{Tab:t6}
\begin{ruledtabular}
\begin{tabular}{cccccc}
 mode&  Case I
 & Case II
   &\cite{Wang:2022ias}  & \cite{Zhao:2022vfr} \\\hline
 $\Omega_{ccc}\to\Xi_{cc} \pi$ & 0.426 & 0.550 & 0.243 & 0.804  \\\hline
 $\Omega_{ccc}\to\Xi_{cc} K $  & 0.0272 & 0.0359   &0.0139 & 0.0673   \\\hline
 $\Omega_{ccc}\to\Omega_{cc} \pi$ & 9.73 & $12.2$  & $11.1 $ &$16.6 $ \\\hline
 $\Omega_{ccc}\to\Omega_{cc}K$  & 0.552 & 0.707  &0.486   & 1.26 \\\hline
 $\Omega_{bbb}\to \Xi_{bb}\pi$    &$1.96\times 10^{-4}$ &$2.71\times 10^{-4}$  & - &$1.44\times 10^{-4}$        \\\hline
 $\Omega_{bbb}\to \Xi_{bb}K$    &$1.57 \times 10^{-5}$  &$2.19\times 10^{-5}$  & - & $1.18\times 10^{-5}$       \\\hline
  $\Omega_{bbb}\to \Xi_{bb}D$    &$2.72\times 10^{-5}$ &$4.44\times 10^{-5}$  & - &$2.01\times 10^{-5}$        \\\hline
 $\Omega_{bbb}\to \Xi_{bb}D_{s}$    &$7.07\times 10^{-4}$  &$1.18\times 10^{-3}$  & - &$5.29\times 10^{-4}$        \
\end{tabular}
\end{ruledtabular}
\end{table}

\section{Summary}
Inspired by the observation of the doubly charmed baryon and great potential on the triply heavy baryon at the LHCb experiments, we study the weak decay of the triply heavy baryon $\Omega_{QQQ}$ to the doubly heavy baryon $\mathcal B_{QQ}$ in this work.
For the $\Omega_{QQQ}$ baryon, we employ the three-quark picture where three identical quarks are treated as independent individuals.
In the decay process, a heavy quark  $Q$ in the initial state decays to a light quark $q$ in the final
state, while the other two quarks $QQ$ are roughly considered as spectators.
The transition matrix elements and  the corresponding form factors are calculated within the LFQM framework.
For the triply heavy baryon $\Omega_{QQQ}$ with three heavy identical quarks, we know very little about its inner structure, i.e. the distance between two $QQ$ is normal or extraordinary, so we make predictions for two cases with different $\beta_{QQ}$, which is a model parameter describing the distance between two heavy quarks.
The form factors are calculated in the space-like region and extended to the time-like region. Using these form factors, some semi-leptonic and non-leptonic decay widths are predicted. We found that the decay rates for Case II are larger than those for Case I for all channels we analyzed here.
Within the quark-diquark picture,
the authors in Refs.~\cite{Wang:2022ias,Zhao:2022vfr} also studied the weak decays of triply heavy baryons in the LFQM. The predictions on $\Gamma (\Omega_{ccc}\to\Xi_{cc}l\bar{\nu}_l)$
and $\Gamma (\Omega_{ccc}\to\Omega_{cc} l\bar{\nu}_l)$ in Ref. \cite{Wang:2022ias} are close to our results in Case I,
but the values of $\Gamma (\Omega_{ccc}\to\Xi_{cc}/\Omega_{cc} l\bar{\nu}_l)$ and $\Gamma (\Omega_{bbb}\to\Omega_{bb} l\bar{\nu}_l)$
in Ref. \cite{Zhao:2022vfr} are significantly different from ours by  an order of magnitude.
For the non-leptonic decay,
our results on $\Gamma (\Omega_{ccc}\to\Omega_{cc} +M)$ in Case I are close to those in Ref. \cite{Wang:2022ias},
but $\Gamma (\Omega_{ccc} \to\Xi_{cc} +M)$ are about twice as large as those in Ref. \cite{Wang:2022ias}.
Compared with the results in Ref.  \cite{Zhao:2022vfr}, our predictions on $\Gamma (\Omega_{ccc}\to\Xi_{cc}+M)$ and $\Gamma (\Omega_{ccc}\to\Omega_{cc}+M)$ in Case I
are about half as large. For the decay $ \Omega_{bbb}\to \Xi_{bb}+M$,
the widths $\Gamma ( \Omega_{bbb}\to \Xi_{bb}+M)$ we obtained in Case I are around 1.5 times larger than those in Ref.  \cite{Zhao:2022vfr}.
Since we employ the three-quark picture instead of the normal diquark picture, different predictions between models can be understood.
We hope that future experiments will measure more data to help us determine which option is more plausible so that we can learn more about the inner structure of  the triply heavy baryon.

\section*{Acknowledgement}

 This work is supported by the National Natural Science Foundation of China (NNSFC)
under the contract No. 12075167, 11975165 and 12235018.

\appendix
\section {The form factors of
 $\Omega_{QQQ}\to \mathcal B_{QQ}$ }

 $\bar {u}_{\xi}(\bar
P,S_z)\gamma_{\mu}\bar P'^\xi \gamma_{5} u(\bar P',S'_z)$ , $\bar
{u}_{\xi}(\bar P,S_z)\bar P'_{\mu}\bar P'^\xi \gamma_{5} u \bar P',S'_z)$,
$\bar {u}_{\xi}(\bar P,S_z) \bar P_{\mu}\bar P'^\xi \gamma_{5} u(\bar
P',S'_z)$, $\bar {u}_{\xi}(\bar P,S_z) g_{\mu}^{\xi} \gamma_{5} u(\bar
P',S'_z)$ are multiplied to the right side of
Eq.(\ref{s1}), and then we have
\begin{eqnarray}\label{s01}
  F_1&=&\int\frac{ d x_2 d^2 k^2_{2\perp}}{2(2\pi)^3}\frac{ d x_3 d^2 k^2_{3\perp}}{2(2\pi)^3}\frac{{\phi^*(x'_i,k'_{i\perp})
  \phi(x_i,k_{i\perp})}Tr[\gamma_{\perp}^\beta(\bar{P'}\!\!\!\!\!\slash+M_0')(p_3\!\!\!\!\!\slash+m_3)
  (\bar{P}\!\!\!\!\slash+M_0)\gamma_{\perp}^\alpha(p_2\!\!\!\!\!\slash-m_2)]}{A} \nonumber \\
   &&\times\sum_{S_z,S'_z}{\rm Tr}[u(\bar{P'},S'_z)\bar u(\bar P',S'_z)
 \gamma_{\perp\beta}\gamma_{5}  (p_1\!\!\!\!\!\slash'+m'_1)\gamma^{\mu}
  (p_1\!\!\!\!\!\slash+m_1) {u}_{\alpha}(\bar P,S_z)\bar{u}_{\xi}(\bar{P},S_z)\gamma_{\mu}\bar P'^\xi \gamma_{5}],\cr
   F_2&=&\int\frac{ d x_2 d^2 k^2_{2\perp}}{2(2\pi)^3}\frac{ d x_3 d^2 k^2_{3\perp}}{2(2\pi)^3}\frac{{\phi^*(x'_i,k'_{i\perp})
  \phi(x_i,k_{i\perp})}Tr[\gamma_{\perp}^\beta(\bar{P'}\!\!\!\!\!\slash+M_0')(p_3\!\!\!\!\!\slash+m_3)
  (\bar{P}\!\!\!\!\slash+M_0)\gamma_{\perp}^\alpha(p_2\!\!\!\!\!\slash-m_2)]}{A} \nonumber \\
   &&\times\sum_{S_z,S'_z}{\rm Tr}[u(\bar{P'},S'_z)\bar u(\bar P',S'_z)
 \gamma_{\perp\beta}\gamma_{5}  (p_1\!\!\!\!\!\slash'+m'_1)\gamma^{\mu}
  (p_1\!\!\!\!\!\slash+m_1) {u}_{\alpha}(\bar P,S_z)\bar{u}_{\xi}(\bar{P},S_z) \bar P'_{\mu}\bar
 P'^\xi \gamma_{5}],\cr
  F_3&=&\int\frac{ d x_2 d^2 k^2_{2\perp}}{2(2\pi)^3}\frac{ d x_3 d^2 k^2_{3\perp}}{2(2\pi)^3}\frac{{\phi^*(x'_i,k'_{i\perp})
  \phi(x_i,k_{i\perp})}Tr[\gamma_{\perp}^\beta(\bar{P'}\!\!\!\!\!\slash+M_0')(p_3\!\!\!\!\!\slash+m_3)
  (\bar{P}\!\!\!\!\slash+M_0)\gamma_{\perp}^\alpha(p_2\!\!\!\!\!\slash-m_2)]}{A} \nonumber \\
   &&\times\sum_{S_z,S'_z}{\rm Tr}[u(\bar{P'},S'_z)\bar u(\bar P',S'_z)
 \gamma_{\perp\beta}\gamma_{5}  (p_1\!\!\!\!\!\slash'+m'_1)\gamma^{\mu}
  (p_1\!\!\!\!\!\slash+m_1) {u}_{\alpha}(\bar P,S_z)\bar{u}_{\xi}(\bar{P},S_z)\bar
P_{\mu}\bar P'^\xi \gamma_{5}],\cr
   F_4&=&\int\frac{ d x_2 d^2 k^2_{2\perp}}{2(2\pi)^3}\frac{ d x_3 d^2 k^2_{3\perp}}{2(2\pi)^3}\frac{{\phi^*(x'_i,k'_{i\perp})
  \phi(x_i,k_{i\perp})}Tr[\gamma_{\perp}^\beta(\bar{P'}\!\!\!\!\!\slash+M_0')(p_3\!\!\!\!\!\slash+m_3)
  (\bar{P}\!\!\!\!\slash+M_0)\gamma_{\perp}^\alpha(p_2\!\!\!\!\!\slash-m_2)]}{A} \nonumber \\
   &&\times\sum_{S_z,S'_z}{\rm Tr}[u(\bar{P'},S'_z)\bar u(\bar P',S'_z)
 \gamma_{\perp\beta}\gamma_{5}  (p_1\!\!\!\!\!\slash'+m'_1)\gamma^{\mu}
  (p_1\!\!\!\!\!\slash+m_1) {u}_{\alpha}(\bar P,S_z)\bar{u}_{\xi}(\bar{P},S_z)g_{\mu}^{
\xi} \gamma_{5}] ,
\end{eqnarray}
with $A=16\sqrt{6x_1x'_1M_0^3M_0'^3(m_1+e_1)(m_2+e_2)(m_3+e_3)(m_1'+e_1')
(m_2'+e_2')(m_3'+e_3')}$.

Simultaneously, $\bar {u}_{\xi}(\bar
P,S_z)\gamma_{\mu}\bar P'^\xi \gamma_{5} u(\bar P',S'_z)$ , $\bar
{u}_{\xi}(\bar P,S_z)\bar P'_{\mu}\bar P'^\xi \gamma_{5} u \bar P',S'_z)$,
$\bar {u}_{\xi}(\bar P,S_z) \bar P_{\mu}\bar P'^\xi \gamma_{5} u(\bar
P',S'_z)$, $\bar {u}_{\xi}(\bar P,S_z) g_{\mu}^{\xi} \gamma_{5} u(\bar
P',S'_z)$  are multiplied to the right side of Eq.(\ref{s2}), one can obtain

\begin{eqnarray}\label{s02}
F_1 &=&{\rm Tr}\{u(\bar{P'},S'_z)\bar u(\bar P',S'_z) \Bigg[\gamma^{\mu} P'^\alpha \frac{f_{1}(q'^{2})}{M}
 + \frac{f_{2}(q'^{2})}{M(M-M')} P'^\alpha P^\mu+
 \frac{f_{3}(q'^{2})}{M(M-M')} P'^\alpha P'^\mu
 \nonumber\\&&+f_4(q'^{2})g^{\alpha \mu}
 \Bigg] \gamma_{5} {u}_{\alpha}(\bar P,S_z)\bar{u}_{\xi}(\bar{P},S_z)\gamma_{\mu}\bar  P'^{\xi}\gamma_{5}\},\cr
 F_2 &=&{\rm Tr}\{u(\bar{P'},S'_z)\bar u(\bar P',S'_z) \Bigg[\gamma^{\mu} P'^\alpha \frac{f_{1}(q'^{2})}{M}
 + \frac{f_{2}(q'^{2})}{M(M-M')} P'^\alpha P^\mu+
 \frac{f_{3}(q'^{2})}{M(M-M')} P'^\alpha P'^\mu
\nonumber\\&&+f_4(q'^{2})g^{\alpha \mu}
 \Bigg]  \gamma_{5} {u}_{\alpha}(\bar P,S_z)\bar{u}_{\xi}(\bar{P},S_z)\bar P'_{\mu}\bar
 P'^{\xi} \gamma_{5}\},\cr
 F_3 &=&{\rm Tr}\{u(\bar{P'},S'_z)\bar u(\bar P',S'_z) \Bigg[\gamma^{\mu} P'^\alpha \frac{f_{1}(q'^{2})}{M}
 + \frac{f_{2}(q'^{2})}{M(M-M')} P'^\alpha P^\mu+
 \frac{f_{3}(q'^{2})}{M(M-M')} P'^\alpha P'^\mu
 \nonumber\\&& +f_4(q'^{2})g^{\alpha \mu}
 \Bigg]   \gamma_{5}  {u}_{\alpha}(\bar P,S_z)\bar{u}_{\xi}(\bar{P},S_z)\bar
P_{\mu}\bar P'^{\xi} \gamma_{5} \},\cr
 F_4 &=&{\rm Tr}\{u(\bar{P'},S'_z)\bar u(\bar P',S'_z) \Bigg[\gamma^{\mu} P'^\alpha \frac{f_{1}(q'^{2})}{M}
 + \frac{f_{2}(q'^{2})}{M(M-M')} P'^\alpha P^\mu+
 \frac{f_{3}(q'^{2})}{M(M-M')} P'^\alpha P'^\mu
 \nonumber\\&& +f_4(q'^{2})g^{\alpha \mu}
\Bigg]   \gamma_{5} {u}_{\alpha}(\bar P,S_z)\bar{u}_{\xi}(\bar{P},S_z)g_{\mu}^{\xi} \gamma_{5}\}.
 \end{eqnarray}

\section{Semi-leptonic decay of  $\Omega_{QQQ}\to \mathcal B_{QQ}  l\bar\nu_l$ }

The helicity amplitudes are given by the form factors
for $\Omega_{QQQ}\to \mathcal B_{QQ} $ \cite{Wang:2022ias}
 \begin{eqnarray}
\label{eq:haad}
H^{V/A}_{1/2,0}&=&\pm\frac{\sqrt{2MM'(w\mp1)}}{2 \sqrt{6 q'^2} M^2}
[2MM'(w\pm1) (M' \mp M) {\cal N}^{V,A}_1(w)-2M(M^2-M'^2+q'^2){\cal N}^{V,A}_4(w)\cr
&&+ \frac{4M^2M'^2}{M-M'} (w^2-1)({\cal N}^{V,A}_2(w)+{\cal N}^{V,A}_3(w))],\cr
H^{V/A}_{1/2,1}&=&- \sqrt \frac{2MM'(w\mp1)}{3}[ {\cal N}^{V,A}_1(w)  \frac{2MM'(w\pm1)}{M^2}+ {\cal N}^{V,A}_4(w)], \cr
H^{V/A}_{{1/2,-1}}&=&-\sqrt{ 2MM'(w\mp1)} {\cal N}^{V,A}_4(w),
\end{eqnarray}
where again the upper (lower)  sign corresponds  to $V(A)$ , ${\cal
  N}^V_i\equiv g_i$, ${\cal N}^A_i\equiv f_i$ ($i=1,2,3,4$) and the
  $q'^2$ is the lepton pair invariant mass with $q'^2=M^2+M'^2-2MM' w$.
The remaining helicity amplitudes  can be obtained using the
relation
$$H^{V,A}_{-\lambda',\,-\lambda_W}=\mp H^{V,A}_{\lambda',\, \lambda_W}.$$
$$H_{\lambda',\,\lambda_W}= H^{V}_{\lambda',\, \lambda_W}- H^{A}_{\lambda',\, \lambda_W}.$$

Partial differential decay rates can be represented in the
following form
\begin{eqnarray}
  \label{eq:darlt}
 \frac{d\Gamma_T}{dw}&=&\frac{G_F^2  |V_{Qq}|^2}{(2\pi)^3}\frac{q'^2
   M'^2\sqrt{w^2-1}}{192 \pi ^3M} [|H_{-1/2,\, 1}|^2+
   |H_{-1/2,\, -1}|^2+|H_{1/2,\, 1}|^2+
   |H_{1/2,\, -1}|^2],\cr
\frac{d\Gamma_L}{dw}&=&\frac{G_F^2 |V_{Qq}|^2}{(2\pi)^3} \frac{q'^2
   M'^2\sqrt{w^2-1}}{192 \pi ^3M} [|H_{1/2,\, 0}|^2+
   |H_{-1/2,\, 0}|^2].
\end{eqnarray}

The differential decay width of the $\Omega_{QQQ}\to \mathcal B_{QQ}
l\bar\nu_l$   can be written as
\begin{eqnarray}
  \label{eq:darlt}
 \frac{d\Gamma}{dw}&=&\frac{d\Gamma_T}{dw}+\frac{d\Gamma_L}{dw}.
\end{eqnarray}

Integrating over the parameter $\omega$, we can obtain the total
decay width
\begin{eqnarray}
  \label{eq:darlt}
\Gamma=\int_1^{\omega_{\rm max}}d\omega\frac{d\Gamma}{dw},
\end{eqnarray}
where $\omega=v\cdot v'$ and the upper bound of the integration
$\omega_{\rm max}=\frac {1}{2}(\frac{M}{M'}+\frac{M'}{M})$ is the
maximal recoil.

The ratio of the longitudinal to transverse decay rates $R$ is
defined by
\begin{eqnarray}
 R=\frac{\Gamma_L}{\Gamma_T}=\frac{\int_1^{\omega_{\rm
     max}}d\omega~\left[ |H_{\frac{1}{2},0}|^2+|H_{-\frac{1}{2},0}|^2
     \right]}{\int_1^{\omega_{\rm max}}d\omega~
     \left[ |H_{-1/2,\, 1}|^2+
   |H_{1/2,\, -1}|^2+|H_{1/2,\, 1}|^2+
   |H_{-1/2,\, -1}|^2\right]}.
\end{eqnarray}


\begin{thebibliography}{99}

\bibitem{LHCb:2017iph}
R.~Aaij \textit{et al.} [LHCb],
Phys. Rev. Lett. \textbf{119}, no.11, 112001 (2017)
doi:10.1103/PhysRevLett.119.112001
[arXiv:1707.01621 [hep-ex]].

\bibitem{LHCb:2018pcs}
R.~Aaij \textit{et al.} [LHCb],
Phys. Rev. Lett. \textbf{121}, no.16, 162002 (2018)
doi:10.1103/PhysRevLett.121.162002 [arXiv:1807.01919 [hep-ex]].


\bibitem{LHCb:2022rpd}
R.~Aaij \textit{et al.} [LHCb],
JHEP \textbf{05}, 038 (2022) doi:10.1007/JHEP05(2022)038
[arXiv:2202.05648 [hep-ex]].

\bibitem{Wang:2017mqp}
W.~Wang, F.~S.~Yu and Z.~X.~Zhao,
Eur. Phys. J. C \textbf{77}, no.11, 781 (2017)
doi:10.1140/epjc/s10052-017-5360-1 [arXiv:1707.02834 [hep-ph]].


\bibitem{Zhao:2018mrg}
Z.~X.~Zhao,
Eur. Phys. J. C \textbf{78}, no.9, 756 (2018)
doi:10.1140/epjc/s10052-018-6213-2
[arXiv:1805.10878 [hep-ph]].


\bibitem{Yu:2017zst}
F.~S.~Yu, H.~Y.~Jiang, R.~H.~Li, C.~D.~L\"u, W.~Wang and Z.~X.~Zhao,
Chin. Phys. C \textbf{42}, no.5, 051001 (2018)
doi:10.1088/1674-1137/42/5/051001
[arXiv:1703.09086 [hep-ph]].

\bibitem{Aliev:2022tvs}
T.~M.~Aliev, S.~Bilmis and M.~Savci,
Eur. Phys. J. C \textbf{82}, no.10, 862 (2022)
doi:10.1140/epjc/s10052-022-10845-5
[arXiv:2206.08253 [hep-ph]].

\bibitem{Hu:2022xzu}
X.~H.~Hu and Y.~J.~Shi,
Phys. Rev. D \textbf{107}, no.3, 036007 (2023)
doi:10.1103/PhysRevD.107.036007
[arXiv:2202.07540 [hep-ph]].

\bibitem{Patel:2008mv}
B.~Patel, A.~Majethiya and P.~C.~Vinodkumar,
Pramana \textbf{72}, 679-688 (2009)
doi:10.1007/s12043-009-0061-4
[arXiv:0808.2880 [hep-ph]].

\bibitem{Brown:2014ena}
Z.~S.~Brown, W.~Detmold, S.~Meinel and K.~Orginos,
Phys. Rev. D \textbf{90}, no.9, 094507 (2014)
doi:10.1103/PhysRevD.90.094507
[arXiv:1409.0497 [hep-lat]].

\bibitem{Yang:2019lsg}
G.~Yang, J.~Ping, P.~G.~Ortega and J.~Segovia,
Chin. Phys. C \textbf{44}, no.2, 023102 (2020)
doi:10.1088/1674-1137/44/2/023102
[arXiv:1904.10166 [hep-ph]].

\bibitem{Chen:2011mb}
Y.~Q.~Chen and S.~Z.~Wu,
JHEP \textbf{08}, 144 (2011)
[erratum: JHEP \textbf{09}, 089 (2011)]
doi:10.1007/JHEP08(2011)144
[arXiv:1106.0193 [hep-ph]].

\bibitem{Faustov:2021qqf}
R.~N.~Faustov and V.~O.~Galkin,
Phys. Rev. D \textbf{105}, no.1, 014013 (2022)
doi:10.1103/PhysRevD.105.014013
[arXiv:2111.07702 [hep-ph]].

\bibitem{Zheng:2010zzc}
W.~Zheng and H.~R.~Pang,
Mod. Phys. Lett. A \textbf{25}, 2077-2088 (2010)
doi:10.1142/S0217732310032962

\bibitem{Jia:2006gw}
Y.~Jia,
JHEP \textbf{10}, 073 (2006)
doi:10.1088/1126-6708/2006/10/073
[arXiv:hep-ph/0607290 [hep-ph]].

\bibitem{Salehi:2022lkt}
N.~Salehi and A.~Abareshi,
Eur. Phys. J. Plus \textbf{137}, no.12, 1298 (2022)
doi:10.1140/epjp/s13360-022-03426-8



\bibitem{Huang:2021jxt}
F.~Huang, J.~Xu and X.~R.~Zhang,
Eur. Phys. J. C \textbf{81}, no.11, 976 (2021)
doi:10.1140/epjc/s10052-021-09729-x
[arXiv:2107.13958 [hep-ph]].

\bibitem{Wang:2018utj}
W.~Wang and J.~Xu,
Phys. Rev. D \textbf{97}, no.9, 093007 (2018)
doi:10.1103/PhysRevD.97.093007
[arXiv:1803.01476 [hep-ph]].

\bibitem{Wang:2022ias}
W.~Wang and Z.~P.~Xing,
Phys. Lett. B \textbf{834}, 137402 (2022)
doi:10.1016/j.physletb.2022.137402
[arXiv:2203.14446 [hep-ph]].

\bibitem{Zhao:2022vfr}
Z.~X.~Zhao,
[arXiv:2204.00759 [hep-ph]].

\bibitem{Geng:2017mxn}
C.~Q.~Geng, Y.~K.~Hsiao, C.~W.~Liu and T.~H.~Tsai,
JHEP \textbf{11}, 147 (2017)
doi:10.1007/JHEP11(2017)147
[arXiv:1709.00808 [hep-ph]].


\bibitem{ODonnell:1995dio}
P.~J.~O'Donnell, Q.~P.~Xu and H.~K.~K.~Tung,
Phys. Rev. D \textbf{52}, 3966-3977 (1995)
doi:10.1103/PhysRevD.52.3966
[arXiv:hep-ph/9503260 [hep-ph]].

\bibitem{Belyaev:1997iu}
V.~M.~Belyaev and M.~B.~Johnson,
Phys. Lett. B \textbf{423}, 379-384 (1998)
doi:10.1016/S0370-2693(98)00152-X
[arXiv:hep-ph/9707329 [hep-ph]].


\bibitem{Choi:1998jd}
H.~M.~Choi and C.~R.~Ji,
Phys. Rev. D \textbf{59}, 034001 (1999)
doi:10.1103/PhysRevD.59.034001
[arXiv:hep-ph/9807500 [hep-ph]].


\bibitem{DeWitt:2003nxf}
M.~A.~DeWitt, H.~M.~Choi and C.~R.~Ji,
AIP Conf. Proc. \textbf{688}, no.1, 79-87 (2003)
doi:10.1063/1.1632196




\bibitem{Cheng:2003sm}
  H.~Y.~Cheng, C.~K.~Chua and C.~W.~Hwang,
  Phys.\ Rev.\  D {\bf 69}, 074025 (2004).



\bibitem{Ke:2011mu}
  H.~W.~Ke and X.~Q.~Li,
  Eur.\ Phys.\ J.\  C {\bf 71}, 1776 (2011)
  [arXiv:1104.3996 [hep-ph]];

\bibitem{Chang:2018zjq}
Q.~Chang, X.~N.~Li, X.~Q.~Li, F.~Su and Y.~D.~Yang,
Phys. Rev. D \textbf{98}, no.11, 114018 (2018)
doi:10.1103/PhysRevD.98.114018
[arXiv:1810.00296 [hep-ph]].

\bibitem{Ke:2007tg}
  H.~W.~Ke, X.~Q.~Li and Z.~T.~Wei,
  Phys.\ Rev.\  D {\bf 77}, 014020 (2008)
  [arXiv:0710.1927 [hep-ph]].

\bibitem{Ke:2012wa}
  H.~W.~Ke, X.~H.~Yuan, X.~Q.~Li, Z.~T.~Wei and Y.~X.~Zhang,
  Phys.\ Rev.\ D {\bf 86}, 114005 (2012)
  doi:10.1103/PhysRevD.86.114005
  [arXiv:1207.3477 [hep-ph]].


 \bibitem{Ke:2017eqo}
  H.~W.~Ke, N.~Hao and X.~Q.~Li,
  J.\ Phys.\ G {\bf 46}, no. 11, 115003 (2019)
  doi:10.1088/1361-6471/ab29a7
  [arXiv:1711.02518 [hep-ph]].

\bibitem{Zhao:2018zcb}
Z.~X.~Zhao,
Chin. Phys. C \textbf{42}, no.9, 093101 (2018)
doi:10.1088/1674-1137/42/9/093101 [arXiv:1803.02292 [hep-ph]].


\bibitem{Guo:2005qa}
  P.~Guo, H.~W.~Ke, X.~Q.~Li, C.~D.~Lu and Y.~M.~Wang,
  Phys.\ Rev.\ D {\bf 75}, 054017 (2007)
  doi:10.1103/PhysRevD.75.054017
  [hep-ph/0501058].


\bibitem{Hu:2020mxk}
X.~H.~Hu, R.~H.~Li and Z.~P.~Xing,
Eur. Phys. J. C \textbf{80}, no.4, 320 (2020)
doi:10.1140/epjc/s10052-020-7851-8
[arXiv:2001.06375 [hep-ph]].


\bibitem{Chua:2019lgb}
C.~K.~Chua,
PoS \textbf{EPS-HEP2019}, 471 (2020)
doi:10.22323/1.364.0471

\bibitem{Chua:2018lfa}
C.~K.~Chua,
Phys. Rev. D \textbf{99}, no.1, 014023 (2019)
doi:10.1103/PhysRevD.99.014023
[arXiv:1811.09265 [hep-ph]].

\bibitem{Ke:2019lcf}
H.~W.~Ke, F.~Lu, X.~H.~Liu and X.~Q.~Li,
Eur. Phys. J. C \textbf{80}, no.2, 140 (2020)
doi:10.1140/epjc/s10052-020-7699-y [arXiv:1912.01435 [hep-ph]].






\bibitem{Ke:2019smy}
  H.~W.~Ke, N.~Hao and X.~Q.~Li,
  Eur.\ Phys.\ J.\ C {\bf 79}, no. 6, 540 (2019)
  doi:10.1140/epjc/s10052-019-7048-1
  [arXiv:1904.05705 [hep-ph]].


\bibitem{Ke:2021pxk}
H.~W.~Ke, Q.~Q.~Kang, X.~H.~Liu and X.~Q.~Li,
Chin. Phys. C \textbf{45}, no.11, 113103 (2021)
doi:10.1088/1674-1137/ac1c66
[arXiv:2106.07013 [hep-ph]].

\bibitem{Li:2021qod}
Y.~S.~Li, X.~Liu and F.~S.~Yu,
Phys. Rev. D \textbf{104}, no.1, 013005 (2021)
doi:10.1103/PhysRevD.104.013005
[arXiv:2104.04962 [hep-ph]].

\bibitem{Li:2021kfb}
Y.~S.~Li and X.~Liu,
Phys. Rev. D \textbf{105}, no.1, 013003 (2022)
doi:10.1103/PhysRevD.105.013003
[arXiv:2112.02481 [hep-ph]].

\bibitem{Li:2022hcn}
Y.~S.~Li and X.~Liu,
Phys. Rev. D \textbf{107}, no.3, 033005 (2023)
doi:10.1103/PhysRevD.107.033005
[arXiv:2212.00300 [hep-ph]].

\bibitem{Geng:2020gjh}
C.~Q.~Geng, C.~W.~Liu and T.~H.~Tsai,
Phys. Rev. D \textbf{103}, no.5, 054018 (2021)
doi:10.1103/PhysRevD.103.054018
[arXiv:2012.04147 [hep-ph]].




\bibitem{Lu:2023rmq}
F.~Lu, H.~W.~Ke, X.~H.~Liu and Y.~L.~Shi,
Eur. Phys. J. C \textbf{83}, no.5, 412 (2023)
doi:10.1140/epjc/s10052-023-11572-1
[arXiv:2303.02946 [hep-ph]].




\bibitem{Cheng:2004ew}
H.~Y.~Cheng and C.~K.~Chua,
JHEP \textbf{11}, 072 (2004)
doi:10.1088/1126-6708/2004/11/072
[arXiv:hep-ph/0406036 [hep-ph]].

\bibitem{Cheng:2004cc}
H.~Y.~Cheng, C.~K.~Chua and C.~W.~Hwang,
Phys. Rev. D \textbf{70}, 034007 (2004)
doi:10.1103/PhysRevD.70.034007
[arXiv:hep-ph/0403232 [hep-ph]].


\bibitem{ParticleDataGroup:2022pth}
R.~L.~Workman \textit{et al.} [Particle Data Group],
PTEP \textbf{2022}, 083C01 (2022)
doi:10.1093/ptep/ptac097.

\end{thebibliography}
\end{document}